\documentclass[prx,aps,twocolumn,superscriptaddress]{revtex4-2}
\usepackage[pdftex]{graphicx}
\usepackage{amsbsy,amssymb,amsmath,bm,mathtools}
\usepackage{xspace}
\usepackage{bm}% bold math
\usepackage{color}
\usepackage{makecell}
\usepackage{url}
\usepackage[section]{placeins}
\usepackage[colorlinks=true,linkcolor=blue,citecolor=blue]{hyperref}
\usepackage{enumitem}

\begin{document}

%\preprint{Preprint}

\title{Machine learning nonequilibrium phase transitions in charge-density wave insulators}

\author{Yunhao Fan}
\affiliation{Department of Physics, University of Virginia, Charlottesville, VA 22904, USA}

\author{Sheng Zhang}
\affiliation{Department of Physics, University of Virginia, Charlottesville, VA 22904, USA}

\author{Gia-Wei Chern}
\affiliation{Department of Physics, University of Virginia, Charlottesville, VA 22904, USA}

\date{\today}

\begin{abstract}
Nonequilibrium electronic forces play a central role in voltage-driven phase transitions but are notoriously expensive to evaluate in dynamical simulations. Here we develop a machine-learning framework for adiabatic lattice dynamics coupled to nonequilibrium electrons, and demonstrate it for a gating-induced insulator-to-metal transition out of a charge-density-wave state in the Holstein model. Although exact electronic forces can be obtained from nonequilibrium Green’s-function (NEGF) calculations, their high computational cost renders long-time dynamical simulations prohibitively expensive. By exploiting the locality of the electronic response, we train a neural network to directly predict instantaneous local electronic forces from the lattice configuration, thereby bypassing repeated NEGF calculations during time evolution. When combined with Brownian dynamics, the resulting machine-learning force field quantitatively reproduces domain-wall motion and nonequilibrium phase-transition dynamics obtained from full NEGF simulations, while achieving orders-of-magnitude gains in computational efficiency. Our results establish direct force learning as an efficient and accurate approach for simulating nonequilibrium lattice dynamics in driven quantum materials.
\end{abstract}

\maketitle

\section{Introduction}

\label{sec:intro}

Machine-learning (ML) methods have emerged as powerful tools for accelerating computational studies in quantum chemistry and materials science by replacing expensive electronic-structure calculations with efficient surrogate models~\cite{snyder2012,brockherde2017,schutt2019,wang2019,tsubaki2020}. A prominent success of this paradigm is the development of ML-based force fields, which enable large-scale quantum molecular dynamics (QMD) simulations~\cite{marx2009} by reproducing atomic forces from first-principles calculations with near--\textit{ab initio} accuracy~\cite{behler2007,bartok2010,li2015,shapeev2016,mcgibbon2017,botu2017,smith2017,zhang2018,behler2016,chmiela2017,chmiela2018,suwa2019,deringer2019,mueller2020,sauceda2020}. These approaches exploit the locality, or nearsightedness, of electronic matter, which ensures that forces acting on a given degree of freedom are predominantly determined by its immediate local environment~\cite{kohn1996,prodan2005}. As a result, complex many-body interactions can be decomposed into site- or atom-resolved contributions that are amenable to efficient learning and scalable simulation.

Most existing ML force-field approaches are energy-based, with forces obtained as derivatives of a learned total energy, as exemplified by the Behler-Parrinello (BP) framework~\cite{behler2007}. In this approach, the total energy is decomposed as $E = \sum_i \epsilon_i$, where each atomic energy $\epsilon_i$ depends only on the local environment of atom $i$~\cite{behler2007,bartok2010}. The intricate many-body dependence of $\epsilon_i$ on its surroundings is approximated by an ML model trained to reproduce both the total energy and the individual atomic forces obtained from quantum-mechanical calculations. Atomic forces are then evaluated as $\mathbf F_i = -{\partial E}/{\partial \mathbf R_i}$, ensuring internal consistency between energies and forces.

An important advantage of BP-type schemes is that symmetry properties can be incorporated naturally. Because the atomic energy $\epsilon_i$ is a scalar and therefore invariant under symmetry operations such as translations and rotations, symmetry-adapted descriptors can be used to enforce the correct transformation properties directly at the level of the ML representation~\cite{behler2007,bartok2010}. Numerous approaches have been proposed for such symmetry-adapted descriptors~\cite{behler2007,bartok2010,li2015,shapeev2016,behler2011,ghiringhelli2015,bartok2013,drautz2019,himanen2020,huo2022}.  Beyond molecular systems, BP-based force fields have been extended to condensed-matter lattice systems, enabling large-scale simulations of adiabatic dynamics in several well-studied lattice Hamiltonians~\cite{ma2019,liu2022,zhang2022,cheng23a,Ghosh24,tyberg25,Jang25}, as well as large-scale Landau-Lifshitz dynamics simulations of quasi-equilibrium correlated electron magnets~\cite{zhang2020,zhang2021,Fan24,cheng23b}. These extensions highlight the flexibility and broad applicability of energy-based ML force-field approaches within equilibrium and near-equilibrium settings.

By contrast, many technologically relevant phenomena—including current-driven ionic motion, voltage-induced phase transitions, electromigration, and spin-transfer torques—involve nonequilibrium electronic states that generate intrinsically nonconservative forces~\cite{lu2012,todorov2010,slonczewski1996,berger1996}. In such regimes, a global potential energy surface generally does not exist, rendering energy-based force constructions inapplicable. Developing ML frameworks capable of accurately learning and predicting nonequilibrium, nonconservative forces therefore represents a major conceptual and practical challenge, and is essential for extending ML-accelerated simulations to driven, open, and far-from-equilibrium quantum systems.

In this work, we demonstrate the feasibility of ML-based direct force learning for nonconservative forces in the context of driven Holstein systems~\cite{holstein1959}. While a general ML representation for nonconservative forces remains an open problem, we exploit a key simplifying feature of the Holstein model: its dynamical degrees of freedom are \emph{scalar} variables representing the amplitudes of local breathing modes. Consequently, the associated electronic forces are also scalar quantities. Invoking the locality principle, we construct a neural-network model that directly predicts these scalar forces from the instantaneous local lattice environment. Importantly, because the output force transforms as a scalar, this approach retains one of the central advantages of BP-type architectures—namely, that the symmetries of the underlying electronic Hamiltonian can be incorporated systematically through symmetry-adapted descriptors, even though the force is not derived from an energy functional.

We apply this ML force-field framework to simulate a voltage-driven insulator-to-metal transition in the Holstein model. At half filling, the equilibrium system is a band insulator characterized by a checkerboard charge-density-wave (CDW) order. When a chemical-potential drop is imposed through coupling to external electrodes, the CDW state becomes unstable, and a transition to a metallic phase is initiated via the nucleation of metallic domains near the electrodes, followed by the propagation of metal-insulator domain walls across the system. The driving forces arising from the nonequilibrium electronic state are computed using real-space nonequilibrium Green’s-function (NEGF) methods~\cite{meir1992,jauho1994,haug2008,diventra2008}. However, NEGF calculations are computationally demanding even for moderate system sizes, and performing them at every time step renders dynamical simulations of nonequilibrium phase transitions prohibitively expensive.

We demonstrate that a BP-type ML model trained on force data generated from NEGF calculations can accurately reproduce both the local electronic forces and the emergent collective dynamics governing the propagation of CDW-metal domain walls. These results establish a general and practical ML framework for direct force learning in nonequilibrium quantum systems, extending ML force-field methodologies beyond near-equilibrium regimes where energy-based constructions and conservative force fields are applicable.

\section{Adiabatic dynamics of Driven Holstein model}

\begin{figure*}[t]
\centering
\includegraphics[width=1.99\columnwidth]{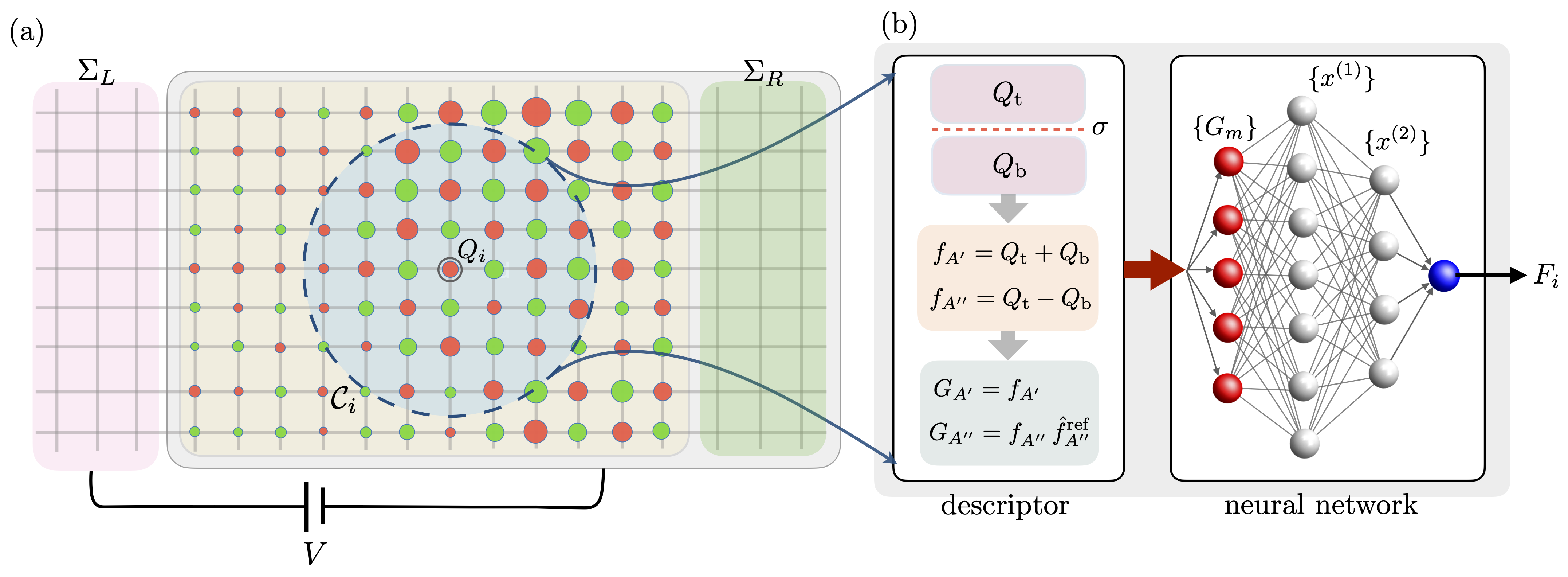}
\caption{Machine-learning framework for adiabatic dynamics of a voltage-driven Holstein system.
(a) Schematic of the gated Holstein setup: a central square-lattice Holstein system is coupled to two electrodes, with the right electrode acting as a substrate and the left electrode providing electrostatic gating. An applied voltage $V$ drives the system out of equilibrium, while local lattice distortions $Q_i$ evolve adiabatically in response to the electronic degrees of freedom. The dashed region indicates the local neighborhood $\mathcal{C}_i$ of a reference site-$i$ used as input to the ML model.  (b) Overview of the ML force-prediction architecture, which generalizes the Behler-Parrinello framework to driven lattice systems. The neighborhood lattice distortions are first decomposed into symmetry-adapted components and processed into descriptor variables that are invariant under the on-site lattice point-group symmetry (here the mirror group $C_s$). These invariant features are then fed into a fully connected neural network, whose output directly yields the local force $F_i$ acting on the lattice degree of freedom at site-$i$. }
    \label{fig:ml-scheme}
\end{figure*}

We begin by formulating nonequilibrium dynamical simulations of a square-lattice Holstein model coupled to two electronic reservoirs and driven by an applied gate voltage. A schematic illustration of the setup is shown in Fig.~\ref{fig:ml-scheme}(a). The system consists of a central interacting region described by the Holstein model, which is connected to noninteracting electronic leads that serve as particle and energy reservoirs. The semi-classical Holstein Hamiltonian~\cite{holstein1959} comprises two components: an effective tight-binding Hamiltonian for the itinerant electrons and a classical elastic energy functional describing the lattice distortions. The electronic part of the Hamiltonian is given by
\begin{equation}
	\label{eq:H1}
	\hat{\mathcal{H}} =-t_{\rm nn}\sum\limits_{\langle ij \rangle}\hat{c}_{i}^{\dagger}\hat{c}^{\,}_{j}
	-g \sum_{i} \hat{n}_i\,Q_i ,   
\end{equation}
Here $\hat{c}_i^\dagger$ ($\hat{c}_i$) creates (annihilates) an electron at lattice site $i$, and $\hat{n}_i = \hat{c}_i^\dagger \hat{c}_i$ is the local electron density operator. The scalar variable $Q_i$ denotes the amplitude of a local collective lattice distortion associated with site $i$, such as the breathing mode of an octahedral cage in perovskite oxides. The first term describes nearest-neighbor electronic hopping on the square lattice with amplitude $t_{\rm nn}$, while the second term encodes a local deformation-potential–type electron–lattice coupling of strength $g$, which energetically favors lattice distortions proportional to the local electronic density. The lattice degrees of freedom are treated classically and are governed by the elastic energy functional
\begin{equation}
	\mathcal{V} =	\frac{k}{2} \sum_i Q_i^2 + \kappa \sum_{\langle ij \rangle} Q_i Q_j .
\end{equation}
where $k$ is an effective on-site elastic constant and $\kappa$ represents a nearest-neighbor antiferrodistortive coupling between lattice distortions. 

The Holstein model provides a paradigmatic description of electron–phonon physics and has been extensively studied in equilibrium in connection with polaron formation, charge ordering, and superconductivity~\cite{bonvca1999,golevz2012,mishchenko2015,costa2018,li2025}. At half filling, the square-lattice Holstein model undergoes a finite-temperature transition into a commensurate CDW phase~\cite{noack1991,zhang2019,chen2019,hohenadler2019}, characterized by a checkerboard modulation of the electronic density: $n_{A/B} = (1 \pm \delta)/2$, where $A$ and $B$ label the two sublattices of the bipartite square lattice and $\delta$ quantifies the CDW amplitude. Through the electron–lattice coupling, this charge modulation is accompanied by a staggered lattice distortion, $Q_{A/B} = \pm \mathcal{Q}$, so that the ordered phase spontaneously breaks the $Z_2$ sublattice symmetry. On symmetry grounds, the CDW transition therefore belongs to the Ising universality class, a conclusion that has been quantitatively confirmed by large-scale quantum Monte Carlo simulations~\cite{noack1991,zhang2019}.

In this work, we focus on the nonequilibrium dynamics of the voltage-driven CDW-to-metal transition. The applied gate voltage drives the electronic subsystem far from thermal equilibrium, destabilizing the CDW order and inducing nontrivial lattice dynamics. Owing to the large mass of the lattice degrees of freedom compared to the electrons, we adopt an adiabatic (Born-Oppenheimer) approximation: the lattice variables ${Q_i}$ evolve dynamically in time, while the electronic subsystem rapidly relaxes to a nonequilibrium steady state determined by the instantaneous lattice configuration. Crucially, unlike conventional Born-Oppenheimer dynamics in equilibrium quantum molecular dynamics, the electronic steady state here is not thermal, but is instead a driven nonequilibrium state set by the bias voltage and reservoir distributions. As discussed below, this nonequilibrium electronic state is computed using the NEGF formalism, which provides the electronic forces acting on the lattice and closes the coupled electron-lattice dynamics.

The time evolution of the lattice degrees of freedom is simulated using over-damped Langevin dynamics, also referred to as Brownian dynamics~\cite{cohen2020,Goetz2022,michielsen1997}, 
\begin{eqnarray}
	\label{eq:BD}
	\frac{dQ_i}{dt} = -\frac{1}{\gamma} F_i + \eta_i(t), 
\end{eqnarray}
where $F_i$ is the driving force, $\gamma$ is an effective damping coefficient and $\eta_i(t)$ denotes a stochastic force with zero mean, $\langle \eta_i(t) \rangle = 0$, and correlations $\langle\eta_i(t)\eta_j(t') \rangle = 2\gamma k_B T \delta_{ij}\delta(t-t')$. 

The total force $F_i$ acting on site $i$ consists of two distinct contributions. The first is an elastic force,  $F_i^{\rm elastic} = -\partial \mathcal{V}/\partial Q_i$ which arises from the classical lattice potential. The second is an electronic force generated by the coupling between the lattice distortions and the driven electronic subsystem. In nonequilibrium steady states, the electronic subsystem cannot, in general, be described by an effective potential energy surface, and the electronic force therefore cannot be written as the gradient of a scalar energy functional. Nevertheless, it has been shown that the force exerted by nonequilibrium electrons on slow classical degrees of freedom can still be evaluated using a generalized Hellmann-Feynman theorem~\cite{diventra2000,todorov2001,lu2012,todorov2010,dundas2009}, $F^{\rm elec}_i = - \langle \partial \hat{\mathcal{H}} / \partial Q_i \rangle$, where the expectation value $\langle \cdots \rangle$ is taken with respect to the quasi-steady, but out-of-equilibrium, electronic state.

Combining the elastic and electronic contributions, the explicit expression for the force is given by
\begin{eqnarray}
	& & F_i = -\biggl\langle \frac{\partial \hat{\mathcal{H}}}{\partial Q_i} \biggr\rangle - \frac{\partial\mathcal{V}}{\partial Q_i} \\
	& & \quad = g \langle \hat{n}_i \rangle - k Q_i - \kappa \sum_{j \in \mathcal{N}(i)} Q_j , \nonumber
\end{eqnarray}
where $\mathcal{N}(i)$ denotes the set of nearest neighbors of site-$i$. While the evaluation of the classical elastic force is straightforward, the electronic contribution—proportional to the local electron density—requires a full nonequilibrium Green’s function (NEGF) calculation and constitutes the dominant computational cost. In a nonequilibrium steady state, the on-site electron density is obtained from the diagonal elements of the lesser Green’s function,
\begin{equation}
	\label{eq:n_i}
	\langle \hat{n}_i \rangle =	\int d\epsilon \,	G^{<}_{ii}\left(\epsilon; \{Q_i\} \right).
\end{equation}
where we explicitly indicate the dependence of $G^<$ on the instantaneous lattice configuration. The lesser Green’s function is determined from the Keldysh equation, 
\begin{eqnarray}
	\label{eq:G_lesser}
    \mathbf G^{<}(\epsilon) = \mathbf G^r(\epsilon) \bm\Sigma^{<}(\epsilon) \mathbf G^a(\epsilon).  \quad
\end{eqnarray}
with $\bm\Sigma^<$ the electrode-induced lesser self-energy. For the present setup, $\bm\Sigma^<$ is diagonal,
\begin{eqnarray}
	\Sigma^{<}_{i  j}(\epsilon) = 2 i \,\delta_{ij} \, \Gamma_{i} \, f_{\rm FD}(\epsilon - \mu_i). 
\end{eqnarray}
with the level-broadening factors $\Gamma_i = \Gamma_{\rm lead}$ or $\Gamma_{\rm bath}$ depending on whether site-$i$ is at the boundaries or in the bulk, and $\mu_i$ the corresponding local chemical potential. As illustrated in Fig.~\ref{fig:ml-scheme}(a), the chemical potential of the substrate (right electrode) and the central region are fixed at the zero $\epsilon_F = \mu_R =0$, which lies at the center of the CDW gap, while the chemical potential of the left electrode is shifted to $\mu_L = - eV$, where $V$ is the applied voltage. 
The retarded and advanced Green’s functions, $\bm G^r$ and $\bm G^a$, respectively, of the central region satisfy $\mathbf G^a = (\mathbf G^r)^\dagger$ and are given by
\begin{eqnarray}
	\label{eq:G_r}
	\mathbf G^r(\epsilon) = (\epsilon \mathbf I - \mathbf H - \bm \Sigma^r)^{-1}, 
\end{eqnarray}
where $H_{ij} =  t_{ij}  - g \delta_{ij} Q_i$ is the tight-binding matrix of the Holstein model and $\bm\Sigma^r$ is the electrode-induced self-energy; more details can be found in Appendix~\ref{app:negf}. Importantly, the energy integral in Eq.~(\ref{eq:n_i}) requires evaluating the Green’s function at many energy points. Each evaluation entails a matrix inversion as in Eq.~(\ref{eq:G_r}), which scales at best as $\mathcal{O}(N^\omega)$ with $\omega \approx 2.37$ when using fast matrix-multiplication algorithms, and as $\mathcal{O}(N^3)$ for standard LU decomposition. As a result, NEGF-based force calculations are computationally demanding and quickly become prohibitive for large-scale dynamical simulations. In the following, we introduce a machine-learning force-field model designed to overcome this computational bottleneck and dramatically improve simulation efficiency.

%The time integration of the Langevin equation is performed using a first-order Euler–Maruyama scheme with time step $\Delta t$. At each time step, the electronic force $F_i^{\rm elec}$ is evaluated either explicitly from NEGF calculations or, in ML-accelerated simulations, predicted directly by the trained force-field model. In the latter case, the electronic force is obtained solely from the instantaneous lattice configuration, eliminating the need for on-the-fly electronic structure calculations and enabling efficient large-scale nonequilibrium lattice dynamics.

\section{Machine Learning force-field Model}

\label{sec:ML}

As discussed in Sec.~\ref{sec:intro}, the feasibility of linear-scaling electronic-structure methods ultimately rests on the locality, or nearsightedness, of many-electron systems~\cite{kohn1996,prodan2005}. Modern machine-learning (ML) techniques provide an explicit and efficient route to embedding this locality principle into practical implementations of $\mathcal{O}(N)$ approaches. Here we propose an ML framework, summarized schematically in Fig.~\ref{fig:ml-scheme}(b), for the direct prediction of the local total force $F_i$ acting on lattice site $i$. Because the forces generated by the driven Holstein model are intrinsically non-conservative, conventional energy-based BP approaches are not applicable. Instead, we train the ML model to directly predict forces. Nevertheless, since the force $F_i$ is a scalar quantity, symmetry-adapted descriptors—namely feature variables that are invariant under the relevant lattice symmetries—can still be employed within the proposed architecture.

The underlying justification for this approach follows directly from the locality principle~\cite{kohn1996,prodan2005}, which implies that the effective force $F_i$ acting on the local lattice mode~$Q_i$ is determined primarily by the structural configuration in its immediate neighborhood. Explicitly, we define the neighborhood centered at site $i$ as the set of lattice distortions within a cutoff radius $r_c$,
\begin{eqnarray}
	\mathcal{C}_i = \left\{ Q_j \, \big| \, |\mathbf r_j - \mathbf r_i | \le r_c \right\}.
\end{eqnarray}
In practice, the local environment $\mathcal{C}_i$ is encoded through a set of feature variables $\bm G = \{G_1, G_2, \cdots \}$ constructed from the structural information within the cutoff region. For a fixed set of parameters in the electronic Hamiltonian, the local force can then be expressed as a universal function of these descriptors,
\begin{eqnarray}
	F_i = \mathcal{F}\!\left( \bm G \right)= \mathcal{F}\bigl( \left\{ G_1(\mathcal{C}_i), G_2(\mathcal{C}_i), \cdots \right\}\bigr).
\end{eqnarray}
The nontrivial dependence of $F_i$ on the local environment is approximated by a fully connected neural network trained on exact NEGF-based force calculations. Conceptually, this procedure yields an effective classical force-field description in the adiabatic limit. At the same time, recent advances in deep-learning architectures provide a systematic and flexible framework for learning the highly nonlinear mapping $\mathcal{F}(\cdot)$ with both high accuracy and computational efficiency, significantly improving the efficiency compared with direct NEGF calculations.

\medskip

\noindent {\bf Symmetry-adapted descriptors}. 
As illustrated schematically in Fig.~\ref{fig:ml-scheme}(b), a central component of the ML force-field framework is the choice of descriptors, which encode the symmetries of the underlying electronic Hamiltonian directly into the learning architecture. Unlike quantum MD simulations, where the relevant symmetry group is the continuous three-dimensional Euclidean group $E(3)$, lattice systems possess only discrete translational symmetry---corresponding to invariance under translations by integer lattice vectors---together with a finite on-site point-group symmetry associated with local rotations and reflections.

Discrete lattice translational symmetry is naturally preserved in our ML framework by applying the same local force-prediction model uniformly to every lattice site. The remaining on-site point-group symmetry can be systematically incorporated through symmetry-adapted descriptors constructed using group-theoretical methods~\cite{ma2019,zhang2022Descriptor}. A number of concrete realizations of such symmetry-aware descriptors and architectures have been developed and successfully applied to lattice models in recent years~\cite{zhang2022,cheng23a,Ghosh24,tyberg25,Jang25,zhang2020,zhang2021,Fan24}.

Although the site symmetry of an ideal square lattice is described by the $D_4$ (or $C_{4v}$) point group, the presence of two electrodes explicitly lowers this symmetry to the simple mirror group $C_s$, whose only nontrivial symmetry operation is a reflection $\sigma$. For a given lattice site $i$, the mirror plane is taken to lie along the $x$ direction and to pass through site $i$. To construct symmetry-invariant feature variables, we first decompose the local lattice configuration $\mathcal{C}_i$ into irreducible representations (IRs) of the point group, which for $C_s$ consist of two one-dimensional IRs, $A'$ and $A''$.
Let $j$ and $\hat{j}$ denote a pair of sites in the neighborhood related by the mirror operation $\sigma$. The corresponding symmetry-adapted basis functions are
\begin{eqnarray}
	f_{A'} = Q_j + Q_{\hat{j}}, \qquad
	f_{A''} = Q_j - Q_{\hat{j}}.
\end{eqnarray}
The symmetric combination $f_{A'}$ is invariant under mirror reflection, whereas the antisymmetric combination $f_{A''}$ changes sign under $\sigma$. While the magnitudes $|f_{A''}|$ could in principle be used as symmetry-invariant features alongside the $A'$ components, this construction discards the $\pm 1$ phase information, in particular the relative phases between different $A''$ basis functions, which encode important geometric information about the local configuration.

To retain this phase information, we introduce the concept of a reference IR, which provides a consistent phase convention for nontrivial IRs. In the present case, a reference for the $A''$ representation is constructed by averaging lattice distortions over extended regions in the neighborhood. Specifically, we define $Q_{\rm t} = \sum_{j \in \mathcal{C}^{\rm t}_i} Q_j$, as the average over the ``top'' region above site $i$, and $Q_{\rm b}$ as the corresponding average over the mirror-related ``bottom'' region; see Fig.~\ref{fig:ml-scheme}(b). These quantities define a reference antisymmetric mode, $f^{\rm ref}_{A''} = Q_{\rm t} - Q_{\rm b}$. Using spatially averaged lattice amplitudes ensures that the reference IR is robust against local fluctuations in $\mathcal{C}_i$.

Importantly, the reference IR allows us to introduce a well-defined phase variable for each antisymmetric basis function,\: $\eta_{A''} = {\rm sign}\bigl( f_{A''} f^{\rm ref}_{A''} \bigr) = \pm 1$. The relative phase between two antisymmetric IR components $f^{(1)}_{A''}$ and $f^{(2)}_{A''}$ is then simply given by $\eta^{(1)}_{A''}\eta^{(2)}_{A''}$. Combining this phase information with the corresponding amplitudes yields the following symmetry-invariant feature variables,
\begin{eqnarray}
	G_{A'} = f_{A'}, \qquad G_{A''} = f_{A''} f^{\rm ref}_{A''}
\end{eqnarray}
Feature variables constructed in this manner from different site pairs in the neighborhood $\mathcal{C}_i$ are used as direct inputs to the neural network, as illustrated in Fig.~\ref{fig:ml-scheme}(b). By construction, this guarantees that the predicted scalar force at each site is invariant under the full set of symmetry operations of the underlying lattice Hamiltonian.

\medskip 
 
\noindent {\bf Neural network implementation.}
As illustrated in Fig.~\ref{fig:ml-scheme}(b), the neural network takes the symmetry-adapted descriptor set $\bm G = { G_{A'}, G_{A''} }$ as input and outputs the predicted local force $F_i$ acting on lattice site $i$. These symmetry-adapted descriptors encode the local structural environment in a form that is invariant under the relevant lattice symmetries. The input dimensionality is determined by the number of such descriptors within a cutoff radius of $r_c = 6$ lattice constants, resulting in a total of 113 input features per lattice site. All neural networks are implemented using the PyTorch framework~\cite{paszke2017,paszke2019}.
The network architecture begins with an embedding layer that maps the 113-dimensional input vector into a 4096-dimensional latent feature space, enabling a high-capacity representation of the local environment. This embedding is followed by seven fully connected hidden layers with progressively decreasing widths, 
\[
	4096 \rightarrow 2048 \rightarrow 1024 \rightarrow 512 \rightarrow 256 \rightarrow 128 \rightarrow 64 \rightarrow 64
\]
and a final single-node output layer that produces the scalar force prediction. Rectified linear unit (ReLU) activation functions are employed throughout all hidden layers~\cite{nair2010}, allowing the network to efficiently capture the strongly nonlinear dependence of the force on the local lattice configuration. A detailed summary of the network architecture, training procedure, and hyperparameter choices is provided in Table~\ref{tab:parameters}.

\medskip

\noindent {\bf Training details and dataset}.
The network parameters are optimized using the Adam optimizer~\cite{kingma2017} with a learning rate of $1 \times 10^{-4}$, by minimizing the mean-squared error (MSE) between the ML-predicted forces and the corresponding reference forces obtained from nonequilibrium Green’s-function (NEGF) calculations,
\begin{eqnarray}
	\mathcal{L} = \sum_i \left| F^{\rm ML}_i - F^{\rm NEGF}_i \right|^2
\end{eqnarray}
While translational symmetry is preserved along the $y$ direction, coupling to external electrodes explicitly breaks translational invariance along the $x$ direction. In this work, we focus on the central region of the square lattice and train the ML model to capture the propagation of the CDW-metal interface during the nonequilibrium transition. The training and validation datasets are generated from NEGF-based dynamical simulations~\cite{zhang2022Gating}, with lattice configurations sampled from different locations within the central part of the lattice, excluding regions within approximately 10 lattice constants of either electrode.  Within this central region, translational invariance is assumed, and a single ML model is expected to provide a faithful description of the domain-wall dynamics across the system. In total, 1096 configurations are used for training, with $30\%$ of the data reserved for validation.

By contrast, distinct ML models are required to describe the near-electrode regions, where different physical processes dominate. In particular, these regions are characterized by (i) the nucleation of metallic domains near the electrode with lower chemical potential during the early stages of the transition, and (ii) the eventual touchdown of the propagating insulator–metal interface at the opposite electrode. Accurately modeling these processes requires force fields trained specifically for the local nonequilibrium environments present near each electrode.

\begin{table}[t]%The best place to locate the table environment is directly after its first reference in text
\begin{ruledtabular}
\begin{tabular}{|c|cc|}
\textrm{Layer}&\textrm{Network}&\\
\colrule
Input (embedding) layer & \makecell[c]{FC(113,4096)\footnote{Fully connected layer with arguments (input size, output size).}\\act\footnote{The activation function.} =ReLU} &\\
\hline
Hidden Layer 1 & \makecell[c]{FC(4096,2048)\\act =ReLU} &\\
\hline
Hidden Layer 2 & \makecell[c]{FC(2048,1024)\\act =ReLU} &\\
\hline
Hidden Layer 3 & \makecell[c]{FC(1024,512)\\act =ReLU} &\\
\hline
Hidden Layer 4 & \makecell[c]{FC(512,256)\\act =ReLU} &\\
\hline
Hidden Layer 5 & \makecell[c]{FC(256,128)\\act =ReLU} &\\
\hline
Hidden Layer 6 & \makecell[c]{FC(128,64)\\act =ReLU} &\\
\hline
Hidden Layer 7 & \makecell[c]{FC(64,64)\\act =ReLU} &\\
\hline
Output Layer & FC(64,1) & \\
\end{tabular}
\end{ruledtabular}
\caption{Fully connected neural network architecture and hyperparameters used for electronic force prediction.}
\label{tab:parameters}
\end{table}

\section{Benchmark and ML-Brownian dynamics simulations}

We first assess the accuracy of the machine-learning (ML) force-field model by benchmarking its predictions for the electronic driving forces against exact nonequilibrium Green’s-function (NEGF) calculations~\cite{zhang2022Gating}. From an ML perspective, this benchmark probes the ability of a direct force-learning approach to generalize across nonequilibrium lattice configurations sampled from different stages of the CDW-to-metal transition, rather than merely interpolating within the training set.

\begin{figure}
\centering
\includegraphics[width=0.99\columnwidth]{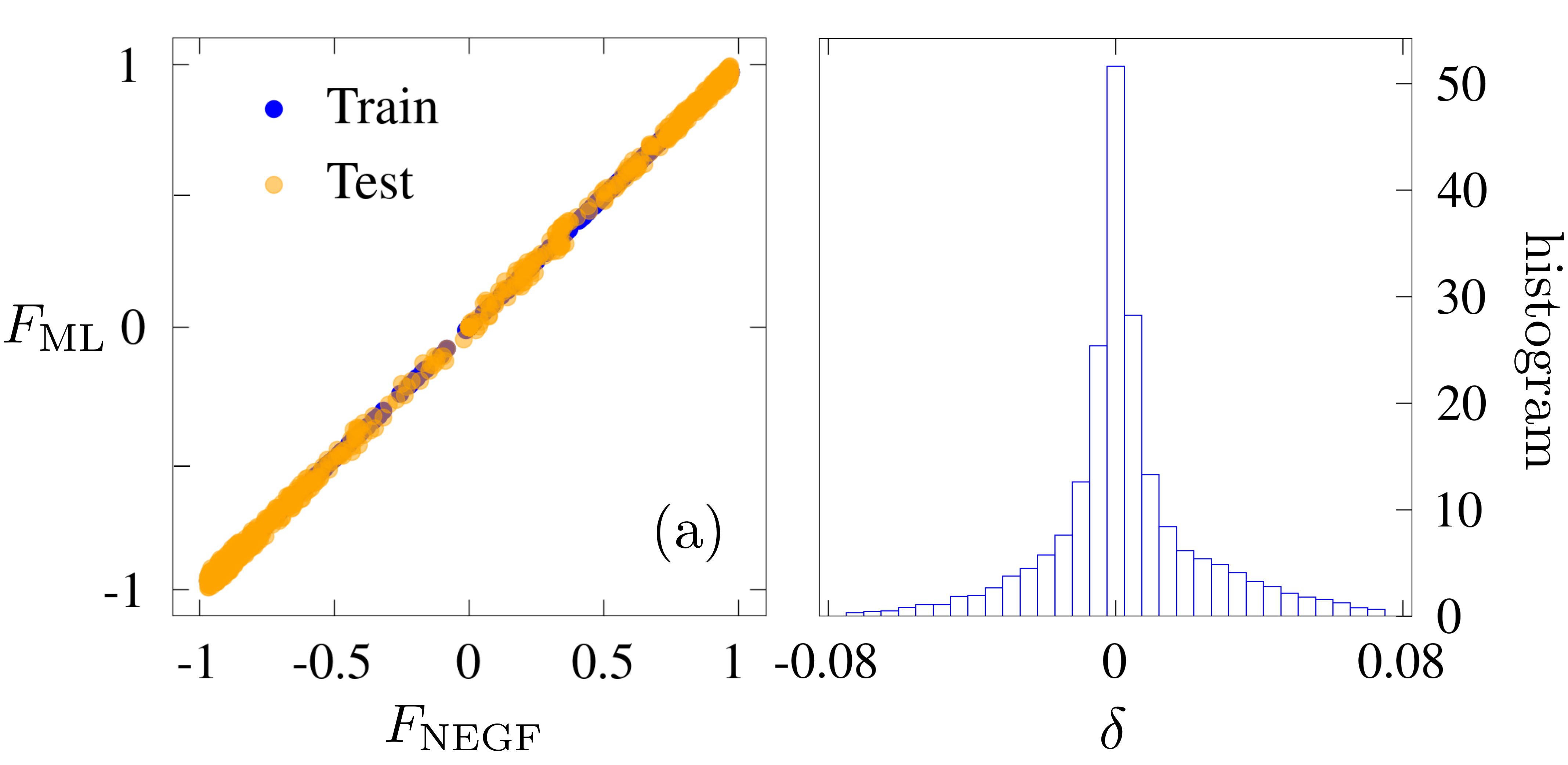}
\caption{Benchmark of ML-predicted electronic forces against NEGF results. (a) Scatter plot comparing ML-predicted forces with reference NEGF forces for the training (blue) and test (orange) datasets. Forces are rescaled to the interval $[-1,1]$ for comparison. (b) Normalized distribution of the prediction error $\delta = F_{\mathrm{NEGF}} - F_{\mathrm{ML}}$ for the test dataset, illustrating the accuracy and bias of the ML force predictions. }
\label{fig:benchmark-force}
\end{figure}

Fig.~\ref{fig:benchmark-force} provides a quantitative comparison between ML-predicted electronic forces and their NEGF counterparts for lattice configurations drawn from nonequilibrium dynamical simulations. As shown in the scatter plot in Fig.~\ref{fig:benchmark-force}(a), the ML predictions closely track the NEGF results over the full range of force values, with both training and test data collapsing tightly onto the diagonal, indicating excellent agreement. This high level of accuracy is reflected in a mean-squared error of $3.0\times10^{-3}$ on the test dataset. The corresponding distribution of prediction errors, shown in Fig.~\ref{fig:benchmark-force}(b), is sharply peaked around zero with a standard deviation of approximately $0.05$, demonstrating the absence of systematic bias and only weak residual fluctuations.

Next, we integrate the trained ML force-field model with the Brownian-dynamics (BD) scheme, using the ML model as an efficient surrogate for explicit nonequilibrium Green's-function (NEGF) calculations in dynamical simulations. At each time step, the driving force entering Eq.~(\ref{eq:BD}) is predicted directly from the instantaneous lattice configuration, thereby eliminating the need for computationally expensive Green's-function evaluations. This ML-BD approach enables large-scale simulations of nonequilibrium lattice dynamics at a fraction of the computational cost of fully self-consistent NEGF calculations. To assess the accuracy of the ML-based dynamics, we directly compare the time evolution of a driven Holstein system obtained using forces computed from explicit NEGF calculations with those predicted by the ML model. Specifically, we consider a Holstein model defined on a $28\times 30$ square lattice, driven out of equilibrium by a chemical-potential difference induced by an external voltage applied across two electrodes.

To provide a dimensionless characterization of the model and simulation parameters, we use the nearest-neighbor hopping amplitude $t_{\rm nn}$ as the fundamental energy scale, which sets the units of energy, electronic broadening, and temperature. Unless otherwise specified, the broadening factors and temperature are fixed at $\Gamma_{\rm lead}=1.0$, $\Gamma_{\rm bath}=0.001$, and $k_{\rm B}T=0.1$, respectively. The electron-lattice coupling strength is quantified by the dimensionless parameter $\lambda = g^{2}/(Wk)$, where $W=8t_{\rm nn}$ is the electronic bandwidth. Simulations are performed with $\lambda=0.28$ and $\kappa/k=0.18$ unless stated otherwise. Time is measured in units of $\gamma^{-1}$, where $\gamma$ is the effective damping coefficient in the Brownian-dynamics scheme, and the integration time step is chosen as $\delta t = 0.1\,\gamma^{-1}$.

As discussed in Sec.~\ref{sec:ML}, the ML force-field model is designed to capture the propagation of the CDW-metal interface during the voltage-induced nonequilibrium phase transition. Lattice configurations in the immediate vicinity of the electrodes are therefore fixed using data obtained from NEGF-BD simulations. These regions correspond to the initial nucleation of metallic domains near one electrode and the final shrinkage of CDW domains near the other. The distinct physical mechanisms underlying these processes are beyond the scope of the present ML framework, which is instead tailored to predict forces associated with the motion of pre-existing domain walls.

\begin{figure}[t]
\centering
\includegraphics[width=0.99\columnwidth]{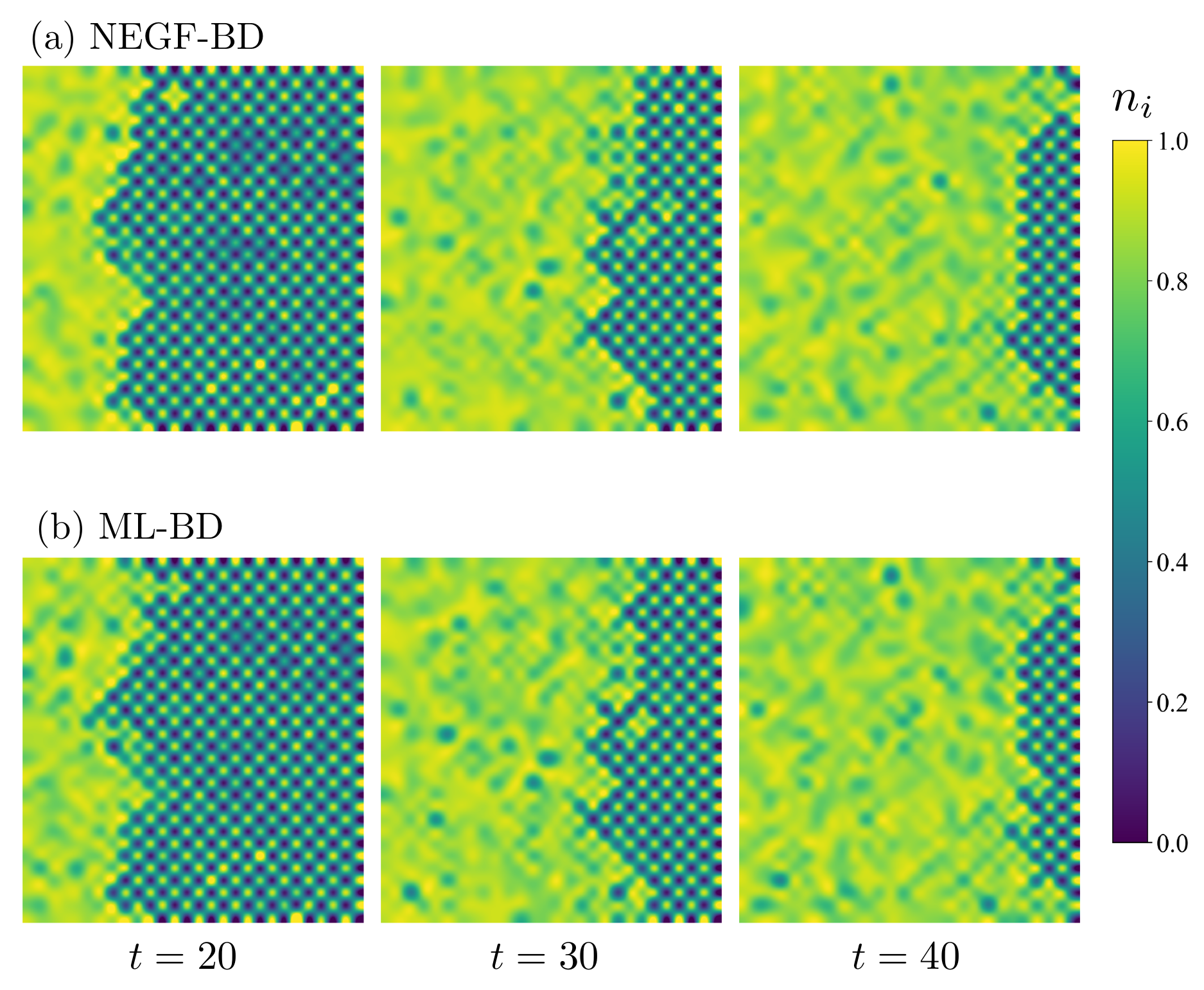}
\caption{Comparison of metallic-phase expansion obtained from NEGF-BD and ML-BD simulations under identical driving conditions. Panels (a) and (b) show snapshots of the on-site electron density $n_i$ on a $28\times30$ lattice region at representative time steps, demonstrating close quantitative and qualitative agreement between the two approaches in capturing the nonequilibrium expansion of the metallic phase.}
\label{fig:snapshots}
\end{figure}

The microscopic mechanism of this voltage-induced transition has been discussed in detail in Ref.~\cite{zhang2022Gating}; here we briefly summarize the essential physics. An applied gate voltage drives the electronic system out of equilibrium and brings the chemical potential $\mu_L$ of the left electrode into resonance with in-gap states localized near the biased lattice edge. This resonant coupling allows electrons to relax into the electrode, suppressing the CDW order and nucleating a metallic layer near the boundary. The transition then proceeds via the propagation of a CDW-metal domain wall into the bulk, with the metallic region advancing layer by layer under intrinsically nonconservative nonequilibrium electronic forces, assisted by thermal fluctuations.

Figures~\ref{fig:snapshots}(a) and (b) show representative snapshots of the local electron density $n_i$ from a NEGF-BD simulation and the corresponding ML-BD simulation, respectively; these configurations are typical of those observed across multiple independent runs. The CDW-to-metal transition is primarily driven by domain-wall propagation, a process that is accurately captured by the ML force-field model. The ML-driven simulations closely reproduce the spatiotemporal expansion of the metallic phase observed in the NEGF-based dynamics. Minor deviations between individual trajectories arise from the combined effects of stochastic noise and residual force-prediction errors. Nevertheless, the close alignment of domain-wall positions and morphologies in the two approaches demonstrates that the ML force field faithfully captures the essential nonequilibrium electronic driving forces governing the CDW-to-metal transition.

\begin{figure}[t]
\centering
\includegraphics[width=0.99\columnwidth]{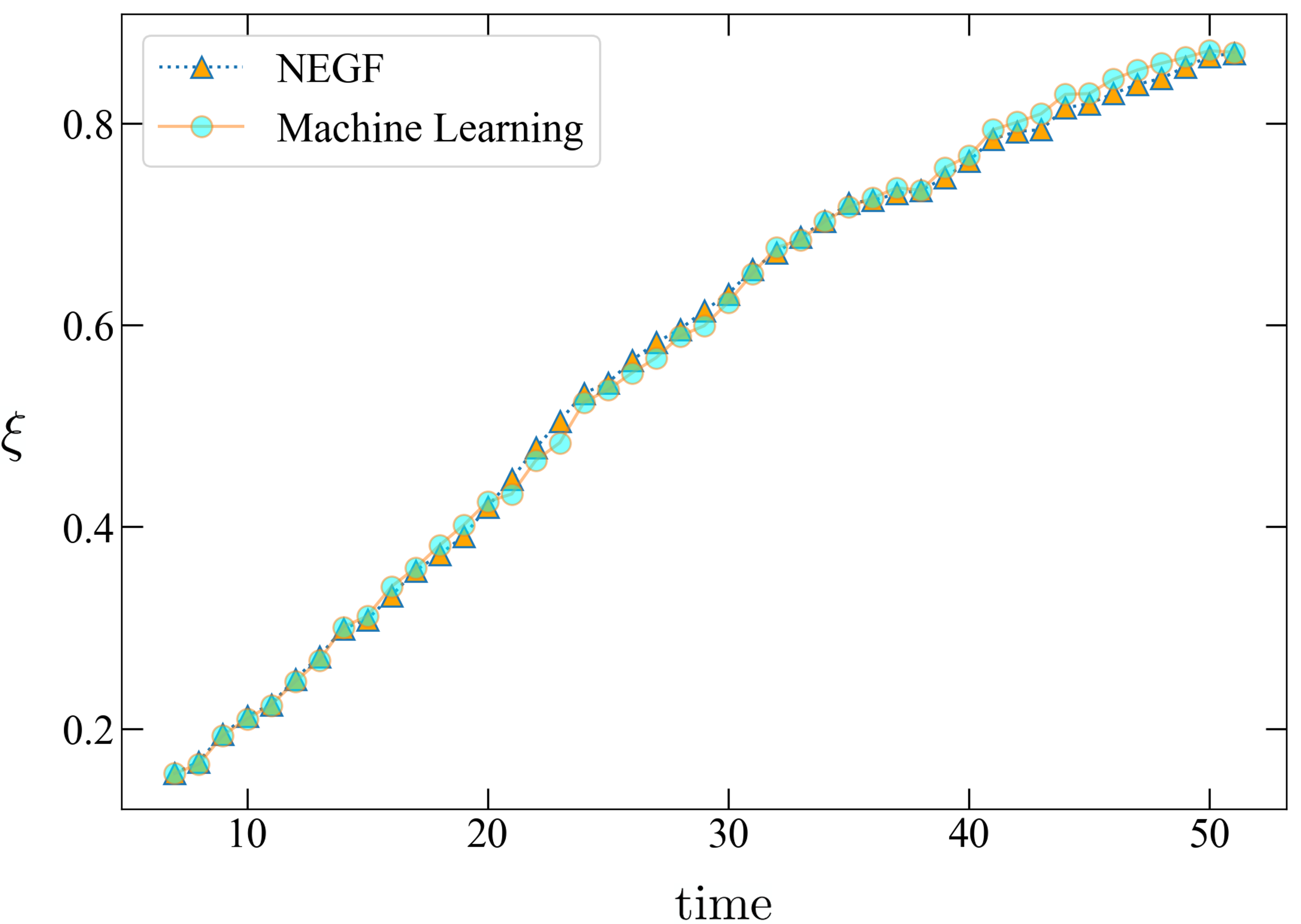}
\caption{Time evolution of the CDW-metal interface position $\xi$ obtained from Brownian-dynamics quench simulations. Yellow triangles represent results from NEGF-driven dynamics, while green circles correspond to ML-driven simulations. The close quantitative agreement between the two trajectories demonstrates that the ML force field accurately reproduces the nonequilibrium electronic driving forces governing domain-wall propagation.}
\label{fig:domain_wall}
\end{figure} 

For a more quantitative comparison, we track the position of the CDW-metal interface, averaged over the transverse $y$ direction, as a function of time. The domain-wall position is defined as the fractional extent of the metallic phase across the lattice. As shown in Fig.~\ref{fig:domain_wall}, the trajectories obtained from ML-BD simulations closely follow those from fully self-consistent NEGF-BD simulations. Although individual realizations display small fluctuations arising from thermal noise inherent in the stochastic dynamics, the overall propagation speed and temporal evolution of the domain wall are in excellent agreement between the two approaches.

This close correspondence demonstrates that residual force-prediction errors in the ML model do not accumulate in a way that qualitatively alters the long-time dynamics. Instead, the ML force field faithfully captures the interplay between nonequilibrium electronic driving forces, lattice elasticity, and thermal fluctuations that governs the kinetics of the voltage-induced phase transition. Importantly, this benchmark goes beyond static force comparisons and establishes that the ML model reliably reproduces emergent, collective nonequilibrium phenomena arising from electron-lattice coupling.

\section{Conclusion and outlook}

\label{sec:conclusion}

In summary, we have demonstrated a scalable machine-learning force-field approach for nonconservative forces generated by nonequilibrium electronic states. Using a voltage-driven Holstein model as a concrete example, we showed that a ML force field constructed from local lattice environments accurately reproduces nonequilibrium electronic forces and the resulting collective lattice dynamics, while enabling simulations that are far more efficient than fully self-consistent NEGF calculations. In contrast to most existing ML-based quantum molecular dynamics methods—which are designed for equilibrium or near-equilibrium electronic systems and rely on conservative, energy-derived forces—our results establish that scalable and local ML force fields can be constructed and deployed even when electronic forces are intrinsically nonconservative and far from equilibrium. This proof-of-principle demonstration shows that scalable ML force-field models are possible not only for equilibrium and near-equilibrium electrons, but also for nonequilibrium electronic states, provided that the underlying electronic response remains sufficiently local.

More generally, constructing ML representations for nonconservative forces requires going beyond conventional Behler–Parrinello–type, energy-based approaches in which forces are obtained as derivatives of a quasi-conservative potential. In certain special cases, such extensions are possible. For example, in spin dynamics, the local driving forces—effective magnetic fields—form a vector field defined on the unit sphere, which allows general nonconservative forces to be represented in terms of two scalar potentials~\cite{zhang2023}. In this setting, the BP framework can be generalized by training an ML model to map the local spin environment onto these two effective energies. However, analogous potential-based representations do not exist for generic nonequilibrium forces in molecular or lattice systems, limiting the applicability of such constructions.

A natural alternative in these cases is to predict forces directly, without introducing intermediate local energies. A central challenge for this strategy is the systematic incorporation of symmetry constraints. One of the appealing features of BP-type force-field models is that symmetries can be enforced at the level of the ML representation through symmetry-invariant descriptors of the local environment. In the present work, we exploit the simplicity of the Holstein model, in which the dynamical lattice degrees of freedom are scalar variables corresponding to the amplitudes of local breathing modes. This allows symmetry-adapted descriptors to be employed even within a direct force-learning framework, enabling us to focus on the essential roles of locality and scalability without additional symmetry-related complications.

Looking ahead, an important open challenge is the extension of scalable ML force-field frameworks for nonequilibrium systems to settings where the relevant forces are vectorial or tensorial, such as Jahn–Teller electron–lattice models~\cite{Ghosh24} or atomic forces in quantum molecular dynamics simulations~\cite{marx2009}. In these cases, direct force prediction must respect nontrivial transformation properties under spatial symmetries, making the construction of symmetry-consistent ML models substantially more demanding than in the scalar-force case considered here. An intriguing and promising direction is to combine direct force learning with equivariant neural networks (ENNs), which enforce the correct covariance of model outputs under symmetry operations by construction. By embedding symmetry constraints at the architectural level rather than through handcrafted descriptors, equivariant models provide a natural framework for learning vector and tensor fields while preserving locality and scalability, making them a compelling candidate for future ML force-field models targeting nonconservative forces in driven, far-from-equilibrium quantum materials.

\appendix

\section{Nonequilibrium Green’s Function (NEGF) calculation}

\label{app:negf}

Here we provide additional details of the nonequilibrium Green’s-function (NEGF) calculations~\cite{meir1992,jauho1994,haug2008,diventra2008} used to generate the training dataset and to benchmark the machine-learning force-field model. The electrode and reservoir degrees of freedom are modeled explicitly by the Hamiltonian
\begin{eqnarray}
	\mathcal{H}_{\rm res} = \sum_{\xi, i} \varepsilon_\xi \, d^\dagger_{i, \xi} d^{\;}_{i, \xi} - \sum_{i, \xi} V_{\xi, i} \bigl(d^\dagger_{i, \xi} c^{\,}_{i} + {\rm h.c.} \bigr). \quad
\end{eqnarray} 
where $d_{i,\xi}$ annihilates a noninteracting fermion in the reservoir (representing either bulk baths or leads), and $\xi$ labels the reservoir quantum numbers, such as band or momentum indices. Sites $i$ located in the bulk are coupled to dissipative baths, while sites at the two open boundaries are coupled to electronic leads.

After integrating out the reservoir fermions, the retarded Green’s-function matrix of the central region takes the form
\begin{eqnarray}
	\mathbf G^r(\epsilon) = (\epsilon \mathbf I - \mathbf H - \bm \Sigma^r)^{-1}, 
\end{eqnarray}
where $H_{ij} = t_{ij} - g \delta_{ij} Q_i$ is the tight-binding matrix of the Holstein model, and
\begin{eqnarray}
	\Sigma^r_{i j}(\epsilon) = \delta_{ij} \sum_\xi  |V_{i,\xi} |^2/ ( \epsilon - \epsilon_\xi + i 0^+)
\end{eqnarray}
is the retarded self-energy induced by the reservoirs. The corresponding level-broadening matrix is defined as $\bm\Gamma = i(\bm\Sigma^{r} - \bm\Sigma^{a})$ and is diagonal, with elements 
\begin{eqnarray}
	\Gamma_{ii} = \pi  \sum_\xi  |V_{i, \xi}|^2 \delta(\epsilon - \epsilon_\xi). 
\end{eqnarray}
For simplicity, we adopt the wide-band limit, assuming a flat reservoir density of states. This approximation renders the broadening parameters energy independent, with two distinct values, $\Gamma_{\rm lead}$ for sites coupled to the electrodes and $\Gamma_{\rm bath}$ for bulk sites.

The lesser Green’s function is obtained using the Keldysh equation,
\begin{eqnarray}
	\mathbf G^{<}(\epsilon) = \mathbf G^r(\epsilon) \bm\Sigma^{<}(\epsilon) \mathbf G^a(\epsilon), 
\end{eqnarray}
where the lesser self-energy is related to the retarded and advanced self-energies through the fluctuation–dissipation relation,
\begin{eqnarray}
	\Sigma^{<}_{i  j}(\epsilon) = 2 i \,\delta_{ij} \, \Gamma_{i} \, f_{\rm FD}(\epsilon - \mu_i). 
\end{eqnarray}
Here $\Gamma_i = \Gamma_{\rm lead}$ or $\Gamma_{\rm bath}$ depending on whether site $i$ is located at the boundaries or in the bulk, and $f_{\rm FD}$ denotes the Fermi–Dirac distribution. The local chemical potential is $\mu_i = \epsilon_F$ for the bulk baths and $\mu_i = \mu_{L/R}$ for the left and right electrodes.

In the setup shown in Fig.~\ref{fig:ml-scheme}(a), the chemical potential of the substrate (right electrode) and the central region is fixed at $\epsilon_F = \mu_R = 0$, which lies at the center of the CDW gap, while the chemical potential of the left electrode is shifted to $\mu_L = -eV$, where $V$ is the applied voltage. Given the lesser Green’s function, the local electron density is obtained from the energy integral of $G^{<}_{ii}(\epsilon)$ [cf. Eq.~(\ref{eq:n_i})], from which the electronic forces acting on the lattice degrees of freedom are evaluated. In practice, the energy integral is computed using a Riemann sum with step size $\Delta \epsilon = 0.003$, corresponding to up to $4\times10^{3}$ energy grid points.

\begin{acknowledgments}
The work is supported by the US Department of Energy Basic Energy Sciences under Contract No. DE-SC0020330. The authors also acknowledge the support of Research Computing at the University of Virginia.
\end{acknowledgments}

\bibliography{refs}

\end{document}